\documentstyle[preprint,tighten,floats,prd,aps]{revtex}
\input epsf

\begin{document}

\draft

\preprint{arch-ive/9610493 \hspace{111mm} UAHEP967}

\title{
   LIGHT GLUINO CONTRIBUTION IN HADRONIC DECAYS \linebreak
                    OF Z BOSON AND $\tau$ LEPTON TO $O(\alpha_s^3)$
}

\author{Louis J.\ Clavelli\footnote{lclavell@ua1vm.ua.edu}
 and Levan R.\ Surguladze\footnote{levan@gluino.ph.ua.edu}}

\address{Department of Physics \& Astronomy, University of Alabama,
                   Tuscaloosa, AL 35487, USA}
\date{August 1996}
\maketitle
\begin{abstract}
The results of calculation of light gluino contributions to
$\Gamma_{Z \rightarrow \mbox{\scriptsize hadrons}}$ and \linebreak
$\Gamma_{\tau^{-} \rightarrow \mbox{\scriptsize hadrons}}$
to $O(\alpha_s^3)$ are presented. The net effect in the case
of Z decay is noticeable. For the $\tau$ width the effect	
is very large and, if a light gluino exists, suggests that
$\alpha_s$ increases by more than 15\% relative to the 
Standard Model analysis. 
\end{abstract}


The existence of the light gluino - a color octet, Majorana fermion,
of mass a few GeV or less, which is the superpartner of the gluon,
is the subject of intensive discussion in the literature
\cite{CLAV,FARR,OTHER}. The possible impact of light gluinos
on LEP/SLC experiments has been discussed recently \cite{FARR,B}. 

One of the best ways to look for a trace of a still undiscovered
light gluino is to evaluate its contributions to the precisely
measured quantities, such as the hadronic decay widths of the Z bozon
and the $\tau$ lepton, using the perturbation theory.
Perturbative QCD (pQCD) calculations of the above quantities up to
$O(\alpha_s^3)$ have been completed in recent years \cite{Rs}-\cite{m^2}.
For a recent review of the status of these calculations see \cite{MOR}.
For the Electroweak contributions see the review \cite{Knirev}.

In the present paper we report results of the evaluation of
pQCD corrections up to 
$O(\alpha_s^3)$ to the above quantities due to the gluino.
We assume that the gluino is light and its mass is anywhere from
less than 1 GeV up to several GeV and the squarks are so heavy that
they decouple for the energy range considered ($\sim M_Z$).
Note however that near the $\tau$ mass, 
the gluino mass has to be well below
$M_{\tau}$, in order our results to be valid for the $\tau$ decay rate.

We start with a brief outline of the theoretical structure of the Z
boson total hadronic decay rate to $O(\alpha_s^3)$
\begin{eqnarray}
\lefteqn{\hspace{-20mm}\Gamma_{Z}
        =\frac{G_FM_Z^3}{8\sqrt{2}\pi}
               \times \biggl\{
           \sum_f \rho_f \biggl(
          v_f^2 \biggl[
             \Gamma_0^V(X_f)
          +\delta_{\mbox{\tiny QED}}^{\mbox{\tiny V}}(\alpha,X_f)
     +\delta_{\mbox{\tiny QCD}}^{\mbox{\tiny V}}
                            (\alpha_s,X_f,X_t,N_{\tilde g})          
                                                \biggr]}\nonumber\\
 && \quad \hspace{17mm}
          +a_f^2\biggl[
               \Gamma_0^A(X_f)
          +\delta_{\mbox{\tiny QED}}^{\mbox{\tiny A}}(\alpha,X_f)
          +\delta_{\mbox{\tiny QCD}}^{\mbox{\tiny A}}
                            (\alpha_s,X_f,X_t,N_{\tilde g})
                                               \biggr]
                                               \biggr)\nonumber\\
 && \quad \hspace{43mm}
     +{\cal L}^{\mbox{\tiny V}}(\alpha_s,X_b,X_t)
     +{\cal L}^{\mbox{\tiny A}}(\alpha_s,X_b,X_t,N_{\tilde g})
                                                   \biggr\}.
\label{Zqq}
\end{eqnarray}
Here the summation index runs over light quark flavors $f=u,d,s,c,b$.
We define $X_f=m_f^2(M_Z)/M_Z^2$ and $X_t=m_t^2(M_Z)/M_Z^2$,
where we use the $\overline{\mbox{\small MS}}$ definition of quark
 masses.
The vector and axial couplings of quark $f$ to the Z boson are
$v_f=2I_f^{(3)}-4e_f\sin^2\theta_{\mbox{\tiny W}}k_f$, 
$a_f=2I_f^{(3)}$. The electroweak self-energy and vertex corrections
are absorbed in the factors $\rho_f$ and $k_f$. The status report
for the electroweak contributions is given in ref.\cite{Knirev}.
The $\Gamma_0^{V/A}(X_f)$ are vector and axial parts of the
well known Born approximation of $\Gamma_Z$ (see, e.g., \cite{Knirev}).
The terms $\delta_{\mbox{\tiny QED}}^{\mbox{\tiny V/A}}(\alpha,X_f)$ 
represent contributions due to diagrams involving at least one photon
exchange between the quarks (see, e.g., \cite{Knirev}).
The QCD contributions are represented by so called nonsinglet
$\delta_{\mbox{\tiny QCD}}^{\mbox{\tiny V/A}}()$ and  singlet
${\cal L}^{\mbox{\tiny V/A}}()$ terms. The singlet part is due to Feynman
graphs with the electroweak currents in separate quark loops mediated
by gluonic states. The other type graphs form the nonsinglet contribution.
$N_{\tilde g}$ in the above equation is the number of light
 gluinos
that can appear virtually in some topological types of graphs, starting
at $O(\alpha_s^2)$. The above QCD terms are calculated up to 
$O(\alpha_s^3)$ within the standard model, 
with no gluinos - $N_{\tilde g}=0$ (see, e.g., \cite{MOR}).
In the present work we calculate the QCD corrections up to the four-loop
level that involve light gluino contributions.

The calculational methods are very similar to that for standard QCD
(for a detailed description see \cite{RMP}). The hadronic decay rate of
the Z boson in the tree level approximation for the electroweak sector
can be evaluated as the imaginary part
\begin{equation}
\Gamma_{Z}
   =-\frac{1}{M_Z}\sum_{f=u,d,s,c,b}
               \mbox{Im}\Pi(m_f,m_t,s+i0)\biggr|_{s=M_Z^2},
\label{ImPi}
\end{equation}
where the function $\Pi$ is defined through a correlation function
of two flavor diagonal quark currents
\begin{equation}
 i\int d^4x e^{iqx}\langle Tj^f_{\mu}(x)j^f_{\nu}(0)\rangle_{0}
   =g_{\mu\nu}\Pi(m_f,m_t,Q^2)-Q_{\mu}Q_{\nu}\Pi'(m_f,m_t,Q^2).
\label{PI}
\end{equation}
Here, $Q^2$ is a large Euclidean momentum $\sim -M_Z^2$. 
The standard neutral weak current of a quark $f$ coupled to 
the Z boson is
$j_{\mu}^{f}=(G_FM_Z^2/2\sqrt{2})^{1/2}
      (v_{f}\overline{q}_f\gamma_{\mu}q_f
      +a_{f}\overline{q}_f\gamma_{\mu}\gamma_5q_f)$.
Because of this structure of the neutral weak current, 
the $\Pi$-function may be decomposed into vector and axial parts
$\Pi(m_f,m_t,Q^2)=\Pi^V(m_f,m_t,Q^2)+\Pi^A(m_f,m_t,Q^2)$.
For further calculational convenience we use the approximation
\begin{equation}
\Pi^{V/A}(m_f,m_t,Q^2)
   =\Pi_0^{V/A}(Q^2,\log m_t)+O(\frac{m_f^2}{Q^2})+O(\frac{Q^2}{m_t^2})
     +\sum_{n\geq 2}\frac{C_n^{V/A}\langle O_n\rangle_{0}}{Q^{2n}}.
\label{Piexpan}
\end{equation}
This is a legitimate expansion since our problem scale is ~$M_Z$.
Thus we work in the limit of zero light quark mass and infinitely
heavy top quark mass.
The last term in the above equation is the nonperturbative
contribution,
parametrized by semi-phenomenological quantities - the so called 
vacuum condensates. $C_n^{V/A}$ are their coefficient functions,
that can also be evaluated perturbatively (see, e.g., \cite{RMP}
and references therein). In this work we ignore
these contributions. The corrections due to light quark masses,
especially for b quark, and finite top mass are not negligible.
They are known up to the order considered (see, e.g., \cite{MOR}
and references therein).
Note that $\Pi_0^{V/A}(Q^2,\log m_t)$ (below we omit the subscript
and superscripts for $\Pi$)  still depends on $m_t$. This is
due to a logarithmic dependence of the axial singlet part on the
top mass that is not suppressed by inverse powers of $m_t$.
Thus the decoupling of heavy particles is not manifest
in the $\mbox{\small MS}$ type prescriptions.
The known mass dependent
corrections can simply be added to our result. On the other hand,
in the limit of vanishing light quark masses, the vector and axial 
parts of the nonsinglet contributions are identical. We calculate first
the nonsinglet vector (axial) part and then we treat the singlet axial
part.
 
The further steps seem to be straightforward.
We write the diagrammatic representation for $\Pi$ and calculate
relevant multiloop Feynman graphs analytically using dimensional
regularization \cite{drg} and the modified minimal subtraction
($\overline{\mbox{\small MS}}$)
prescription \cite{MSB}. After the renormalization of coupling we can
get the final result using eq.(\ref{ImPi}). Unfortunately this
straightforward scheme is realistic only up to three-loop level, since
even the advanced computer program HEPLoops \cite{HEPL}
for evaluation of multiloop Feynman graphs can evaluate
diagrams only up to three-loops.
Fortunately, with the aid of the renormalization group and a 
unique feature of the  $\overline{\mbox{\small MS}}$ prescription one
can reduce the four-loop calculation to the evaluation of only one-,
two-, and three-loop graphs. The detailed outline of the method
and further references are given in ref.\cite{RMP}.

The running strong coupling obeys the renormalization group equation
$\alpha_s\beta(\alpha_s)=\mu^2 d\alpha_s/d\mu^2$,
where the QCD $\beta$ function coefficients for $\mbox{\small MS}$ type
schemes are defined as follows
$\beta(\alpha_{s})=-\beta_{0}(\alpha_s/4\pi)
-\beta_1(\alpha_s/4\pi)^2+O(\alpha_s^3)$, 
$\beta_0=11C_A/3-4TN/3-2C_A N_{\tilde g}/3$,
$\beta_1=34C_A^2/3-20C_ATN/3
-4C_FTN-16C_A^2 N_{\tilde g}/3$.
In the above coefficients we include light gluino contributions.
The corresponding terms are obtained from the analysis of all diagrams
contributing to one-  and two-loop $\beta$-function coefficients
and known results \cite{A} are confirmed.

The evaluation of the three-loop approximation to the decay rate
including the light gluino effect is fairly trivial, 
since there are only two three-loop graphs where the gluino can appear.
We use the three-loop graph-by-graph results of \cite{Rs}
and have confirmed three-loop numerical results given in \cite{B}.
Note that there are no
gluino contributions at one or two-loop levels. At the four-loop level,
the situation is more complex. Although we use still unpublished
four-loop  diagrammatic results from the first work in \cite{Rs},
it was necessary to reanalyze all 27 four-loop graphs
where the light gluino can appear.  These graphs contribute
with different color weights in the case of gluino.
There are several graphs that required recalculation with the
help of the HEPLoops program \cite{HEPL}. Two of those graphs
are shown in Fig.1a and Fig.1b.
\begin{figure}
\hskip 4.0cm
\epsfxsize=3in \epsfysize=1in \epsfbox{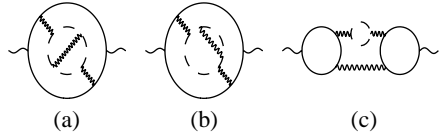}
\caption{Four-loop Feynman graphs with light gluino contributing
in the nonsinglet (a,b) and singlet (c) parts of $\Gamma_Z$.
The dashed lines correspond to light gluino propagators.
The solid and wave lines correspond to quark and gluon propagators.}
\label{Fig.1}
\end{figure}
In the previous standard QCD calculations \cite{Rs},
contributions from those and several other graphs were taken into
account with the help of the full two-loop gluon propagator inserted
into a two-loop graph. In the case of a light gluino it was necessary
to treat these graphs individually, because of different
color and symmetry weights.

For the nonsinglet part of the four-loop QCD correction to the Z decay
rate, including the light gluino contribution, we obtained the
following  $\overline{\mbox{\small MS}}$ analytical result
\begin{eqnarray}
\lefteqn{
  \delta_{\mbox{\tiny QCD}}^{\mbox{\tiny V/A}}
                     (\alpha_s,N_{\tilde g})=
         \biggl(\frac{\alpha_s(M_Z)}{4\pi}\biggr)(3C_F)}  \nonumber\\
  && \quad     +\biggl(\frac{\alpha_s(M_Z)}{4\pi}\biggr)^2
          \biggl\{C_F^2\biggl(-\frac{3}{2}\biggr)
             +C_FC_A\biggl[\frac{123}{2}-44\zeta(3)
                           -(11-8\zeta(3))N_{\tilde g} \biggr]
                                                      \nonumber\\
  && \quad   \hspace{109mm}
                       -N_fTC_F(22-16\zeta(3))\biggr\}
                                                       \nonumber\\
  && \quad   +\biggl(\frac{\alpha_s(M_Z)}{4\pi}\biggr)^3 
          \biggl\{C_F^3\biggl(-\frac{69}{2}\biggr)     \nonumber\\
  && \quad   \hspace{17mm}
    -C_F^2C_A\biggl[127+572\zeta(3)-880\zeta(5)
      -(36+104\zeta(3)-160\zeta(5))N_{\tilde g} \biggr]
                                                        \nonumber\\ 
  && \quad  \hspace{17mm}
           +C_FC_A^2\biggl[\frac{90445}{54}-\frac{10948}{9}\zeta(3)
                        -\frac{440}{3}\zeta(5)
       -\biggl(\frac{33767}{54}-\frac{4016}{9}\zeta(3)
                        -\frac{80}{3}\zeta(5) \biggr)N_{\tilde g}
                                                                \nonumber\\
  &&  \quad \hspace{109mm}                         
       +\biggl(\frac{1208}{27}-\frac{304}{9}\zeta(3) \biggr)
                                                   N_{\tilde g}^2
                                             \biggr]             \nonumber\\
  && \quad \hspace{17mm}        
           -N_fTC_F^2[29-304\zeta(3)+320\zeta(5)]                \nonumber\\
  && \quad \hspace{17mm}          
           -N_fTC_FC_A\biggl[\frac{31040}{27}-\frac{7168}{9}\zeta(3)
                                 -\frac{160}{3}\zeta(5)
           -\biggl( \frac{4832}{27}-\frac{1216}{9}\zeta(3) \biggr)
                                                 N_{\tilde g}
                                                       \biggr]    \nonumber\\ 
  && \quad \hspace{17mm}       
           +N_f^2T^2C_F\biggl[\frac{4832}{27}
                                 -\frac{1216}{9}\zeta(3)\biggr]
              -\pi^2C_F\biggl[\biggl(\frac{11}{3}C_A
              -\frac{4}{3}N_fT\biggr)
         -\frac{2}{3}C_A N_{\tilde g}\biggr]^2\biggr\}.
\label{eq:Ranalytic0}
\end{eqnarray}
In the above expression the logarithmic contributions are summed up 
into the running constant by taking $\mu^2=M_Z^2$. Those contributions
can trivially be restored using the renormalization group (see \cite{RMP}).
Inserting the standard SU$_{\mbox{\scriptsize c}}$(3) eigenvalues
of the Casimir operators for the fundamental and adjoint
representations $C_F=4/3$, $C_A=3$ and also T=1/2, we obtain the
following numerical result
\begin{eqnarray}
\lefteqn{ \hspace{-17mm}
  \delta_{\mbox{\tiny QCD}}^{\mbox{\tiny V/A}}
                     (\alpha_s,N_{\tilde g})
   =\frac{\alpha_s(M_Z)}{\pi}
   +\biggl(\frac{\alpha_s(M_Z)}{\pi}\biggr)^2
                                   (1.9857-0.1153N_f-0.3459N_{\tilde g})}
                                                              \nonumber\\
 && \quad \hspace{9mm}                                       
        +\biggl(\frac{\alpha_s(M_Z)}{\pi}\biggr)^3
        [-6.6369-1.2001N_f-0.0052N_f^2 \nonumber\\
 && \quad \hspace{39mm}
                     -2.8505N_{\tilde g}
                               -0.0311N_fN_{\tilde g}
             -0.0466N_{\tilde g}^2)\biggr].      
\label{eq:Rnum}
\end{eqnarray}

For the singlet part the light gluino contribution to $O(\alpha_s^3)$
shows up in only one single graph (Fig.1c) in the axial channel.
In the vector channel,
the corresponding graph vanishes due to Furry's theorem \cite{Furry}.
The nonvanishing of the graphs like in Fig.1c is due to a large mass
splitting in the top-bottom doublet. This effect was first evaluated
exactly in the 
three-loop level \cite{Kni} and then extended to
four-loop using the large mass expansion method \cite{Lar}. We use
the result of \cite{Lar} to extract the light gluino contribution
to the four-loop axial singlet part. We obtain 

\begin{eqnarray}
\lefteqn{ \hspace{-27mm}
       {\cal L}^{\mbox{\tiny A}}(\alpha_s,X_t,N_{\tilde g})
   =-\biggl(\frac{\alpha_s(M_Z)}{\pi}\biggr)^2
     \biggl[\frac{37}{12}+\log X_t
                            +O(X_t^{-1})     \biggr]} \nonumber\\
 && \quad \hspace{-7mm}
    -\biggl(\frac{\alpha_s(M_Z)}{\pi}\biggr)^3
     \biggl[\frac{6401}{216}-\zeta(3)+\frac{7}{6}\log X_t
               -\frac{11}{4}\log^2 X_t \nonumber\\
 && \quad \hspace{23mm}
       -N_f\biggl(
            \frac{25}{36}-\frac{1}{9}\log X_t
               -\frac{1}{6}\log^2 X_t \biggr) \nonumber\\
 && \quad \hspace{23mm}
       -N_{\tilde g}\biggl(
            \frac{25}{12}-\frac{1}{3}\log X_t
               -\frac{1}{2}\log^2 X_t \biggr) \nonumber\\
 && \quad \hspace{23mm}
       -\frac{\pi^2}{3}\biggl(\frac{11}{4}-\frac{1}{6}N_f
               -\frac{1}{2}N_{\tilde g}\biggr) 
         +O(X_t^{-1})\biggr].
\label{LA}
\end{eqnarray}
Note that throughout this paper the strong coupling is defined for
five flavor effective theory.

For completeness, we also give the result for singlet vector part
in the limit of vanishing light quark masses and infinitely heavy
top mass \cite{Rs}, although the gluino does not contribute here
\begin{equation}
 {\cal L}^{\mbox{\tiny V}}(\alpha_s)
   =-\biggl(\frac{\alpha_s(M_Z)}{\pi}\biggr)^3 (0.4132 +O(X_t^{-1}))
 (\sum_{f=u,d,s,c,b}v_f)^2.
\label{LVnum}
\end{equation}
The terms of $O(X_t^{-1})$ \cite{Lar} are about two orders of magnitude
less than the leading term \cite{Rs} and are completely negligible.

Next, we use our result to calculate the light gluino contribution
to the $\tau$-lepton decay rate to $O(\alpha_s^3)$ in perturbative QCD.
We consider the familiar ratio
$R_{\tau}=\Gamma(\tau^{-}\rightarrow\nu_{\tau}+\mbox{hadrons})/
      \Gamma(\tau^{-}\rightarrow\nu_{\tau}e^{-}\overline{\nu}_{e})$.
Here we are interested only in perturbation theory contributions and
we do not consider nonperturbative and instanton corrections.
For the calculational method and references see \cite{RMP}.
As before, we work
in the limit of vanishing light quark masses and infinitely large
heavy quark masses. Note that for the scale
$\sim M_{\tau}$, u,d,s quarks are considered as light and c,b,t are
heavy quarks.
We use our diagrammatic results obtained for the Z boson case
to evaluate correlator of the charged weak currents of quarks
coupled to W boson.
We obtain the following result for QCD perturbative contributions to 
$R_{\tau}$, including a light gluino

\begin{eqnarray}
\lefteqn{\hspace{-12mm}R_{\tau}^{\mbox{\scriptsize{pert}}}(M_{\tau}^2)= 
   3(0.998\pm 0.002)\biggl\{ 1
        +\frac{\alpha_s(M_{\tau}^2)}{\pi}
        +\biggl(\frac{\alpha_s(M_{\tau}^2)}{\pi}\biggr)^2
          \biggl[\frac{769}{48}-9\zeta(3)
        -N_{\tilde g}\left(\frac{85}{24}-2\zeta(3)\right)\biggr]}
                                                            \nonumber\\
  && \quad 
           +\biggl(\frac{\alpha_s(M_{\tau}^2)}{\pi}\biggr)^3
   \biggl[\frac{363247}{1152}-\frac{81}{8}\zeta(2)-\frac{2071}{8}\zeta(3)
        +\frac{75}{2}\zeta(5) \nonumber\\
 && \quad 
            -N_{\tilde g}
          \biggl(\frac{10787}{72}-\frac{9}{2}\zeta(2)
             -\frac{649}{6}\zeta(3)+\frac{25}{3}\zeta(5)\biggr)
        +N_{\tilde g}^2
         \biggl(\frac{3935}{288}-\frac{1}{2}\zeta(2)
             -\frac{19}{2}\zeta(3)\biggr) \biggr] \nonumber\\
 && \quad \hspace{77mm}
       +O(\frac{M_{\tau}^2}{m_c^2})
       +O(\frac{m_{\tilde g}^2}{M_{\tau}^2}) \biggr\},
\label{eq:Rtauanalytic1}
\end{eqnarray}
and numerically we get
\begin{eqnarray}
\lefteqn{R_{\tau}(M_{\tau}^2)=3(0.998\pm 0.002)\biggl[ 1
        +\frac{\alpha_s(M_{\tau})}{\pi}
        +\left(\frac{\alpha_s(M_{\tau})}{\pi}\right)^2
              (5.2023-1.1376N_{\tilde g}) } \nonumber\\
 && \quad \hspace{7mm}
        +\left(\frac{\alpha_s(M_{\tau})}{\pi}\right)^3
          (26.3659-21.0358 N_{\tilde g}+1.4212N_{\tilde g}^2)
          +O(\frac{M_{\tau}^2}{m_c^2})
          +O(\frac{m_{\tilde g}^2}{M_{\tau}^2}) \biggr].
\label{eq:Rtaunumer}
\end{eqnarray}
The three- and four-loop corrections due to a light gluino
are very large and they increase the extracted $\alpha_s(M_{\tau})$ by
more than 15\%. (On the extraction of $\alpha_s(M_{\tau})$ from
the $\tau$-lepton decay width see, e.g., \cite{Pich} and references
therein.)
For the case $N_{\tilde g}=0$
Eqs.\ (\ref{eq:Rtauanalytic1}) and (\ref{eq:Rtaunumer}) agree with the
known results \cite{Rs}.

Summarizing, we have calculated the light gluino contribution to
hadronic decay rates of the Z boson and the $\tau$-lepton to four-loop
level in perturbative QCD. The corrections in the case of the Z boson 
decrease the three- and four-loop coefficients by about 25\% each.
The net effect of a light gluino is to increase $\alpha_s(M_Z)$
by about 2\%. In the case of the $\tau$-lepton, the corrections are
very large and they decrease the
three- and four-loop coefficients by 22\% and 74\% respectively.
The $\tau$ hadronic decay rate remains a major counter-indication
to the hypothesis of a light gluino. The value of $\alpha_s(M_{\tau})$
extracted from $\tau$ decay with a light gluino and extrapolated to
$M_Z$ overestimates the direct measurement of $\alpha_s(M_Z)$.
Therefore, if a light gluino exists there must be appreciable, positive
contributions to the $\tau$ hadronic width from, for instance, the
nonperturbative region or other as-yet-unknown sources.

\acknowledgements

We thank P.\ W.\ Coulter and G.\ Farrar for beneficial conversations.
This work was supported by the U.S. Department of Energy under
grant No.\ DE-FG02-96ER-40967.

\vspace{7mm}

\noindent
{\bf Note Added} After this paper was completed, we received a paper 
\cite{chetnew}, where the light gluino contribution was calculated
to four-loops for the vector part of the Z boson decay rate.
The result is in complete agreement with our eq.(\ref{eq:Rnum}).

\end{document}